# Green Time-Critical Fog Communication and Computing

Hanna Bogucka[†], Bartosz Kopras[†], Filip Idzikowski[†], Bartosz Bossy[†], Paweł Kryszkiewicz[†]

[†]Poznan University of Technology, Poland

**Abstract**

Fog computing allows computationally-heavy problems with tight time constraints to be solved even if end devices have limited computational resources and latency induced by cloud computing is too high. How can energy consumed by fog computing be saved while obeying latency constraints and considering not only computations but also transmission through wireless and wired links? This work examines the latency and energy consumption sources in fog networks and discusses models describing these costs for various technologies. Next, resource allocation strategies are discussed considering the various degrees of freedom available in such a complex system, and their influence on energy consumption and latency. Finally, a vision for a future distributed, AI-driven resources allocation strategy is presented and justified.

## I. Introduction

Together with the rapid development of modern Information and Communication Technologies (ICT), the energy consumption of these technologies increases. Although the ICT sector's emissions are predicted to stabilize at 1.25 Gt$CO_2$e in 2030 [1], the energy cost of communication and computing services is continuously subject to minimization by the service providers and consumers. This is why energy efficiency is a key paradigm for modern contemporary and future networks, including the Fifth Generation (5G) and Sixth Generation (6G) systems. These networks and services involve both Communication and Computing (2C) of information across the network, and thus, 2C services should be handled (optimized) jointly.

The idea of fog or edge computing is proposed for 5G/6G communication systems and future ICT networks [2]. This technology is essentially a hierarchical, balanced network organization where communication and computing tasks can be performed flexibly using diverse resources available in a network. Fog is an architecture that distributes communication and computation services along the cloud-to-things continuum [2]. It includes information processing, storage, control, and networking to serve many growing applications. A representative instance of the fog network is shown in Fig. 1. Things, such as cars, cellphones, and other linked devices, are present in the things tier. Powerful data servers are deployed in the cloud layer. Connected computing devices (PCs, servers, computing clusters, etc.) that can process, communicate, and store data are located in the fog tier. Multiple hierarchical levels may exist in the fog tier. Collaboration including both vertical and horizontal communication is possible between them.

The execution of a task can be assigned to a (near or distant) fog node, the cloud, or carried out locally, depending on the Quality of Service (QoS) metrics that need to be guaranteed for that task. Information flow is depicted in Fig. 1 for a few examples of use cases, including vehicular communication, remote control in industrial or medical settings (usually Ultra-Reliable, Low-Latency Communication (URLLC)), task offloading from a device with low processing power and memory, content cashing (typically enhanced Mobile BroadBand communication (eMBB)), or telemetry data flow (usually massive Machine-Type Communication (mMTC)).

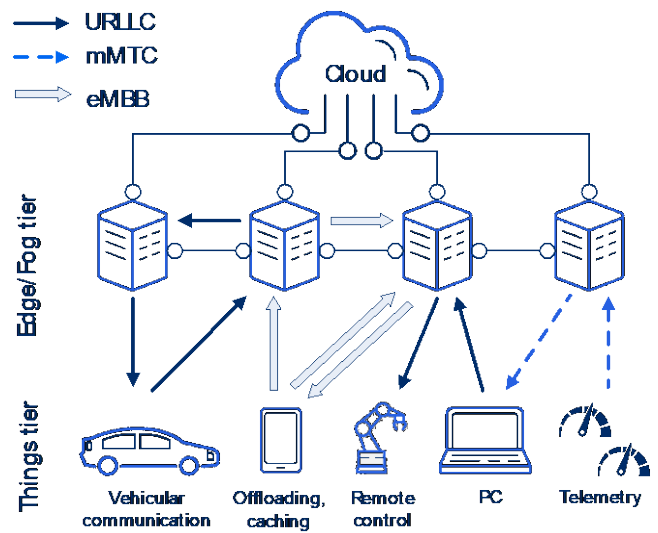

Figure 1. 2C fog network and its optimization platform.

The objective of this work is to improve the energy efficiency of fog networks for mission-critical applications, i.e., those constrained by the deadline of task execution. We (i) jointly optimize resource allocation for communication and computing, (ii) consider various task allocation schemes, (iii) consider the adaptation of clock fre-quency and packet generation rate, (iv) discuss resource allocation with Artificial Intelligence (AI). In Section II, we provide an overview of the causes (devices and processes) of energy consumption in wireless and wired parts of a network and computing machines. Section III presents options for energy consumption minimization with latency and Age of Information (AoI) constraints as well as representative use cases and optimization results. In Section IV, we discuss AI-based practical methods for reducing the energy consumption of a fog network. We conclude our work in Section V.

## II. Key Devices and Processes Affecting Energy Consumption and Latency in Fog Networks

The decision of where to process a computing task from the perspective of energy consumption and latency is affected by the performance of: the wireless part of the network, the wired part of the network, and the devices performing computations themselves.

### A. Wireless access networks

Energy consumption depends on numerous factors, including the number of bits to be transmitted, required bitrate, transmission channel properties, e.g., path loss, fading, and the utilized wireless transmission standard, e.g., 5G or WiFi with its configuration. The most common approach to the modeling (limited in its application range) is to perform measurements of a wireless transceiver under various conditions, e.g., payload and path loss, and extrapolate the values to cover other use cases as well. The main drawback of this approach is characterization limited to a single device. The other approach is to characterize every single element of a wireless transceiver in terms of its energy consumption, e.g., Analog-Digital Converter (ADC) and coder. However, this results in a multi-parameter model which is difficult to be configured to resemble real products. While [3] can be used as a first reference point for WiFi devices, [4] shows energy consumption for a 4G/5G smartphone, and [5] for an LTE network. It is visible in Table I that the representative energy efficiency of a WiFi modem equals around 39-45 $nJ/b$ at each side of a wireless link (assumed path loss of 83 dB). While the 5G transmission is about 10 times less efficient from a 5G terminal perspective, it still outperforms an LTE terminal. However, the energy efficiency of an LTE BS is significantly lower, resulting in 45 $\mu J/b$ on average [5]. As all these numbers were obtained in different environments, under different test conditions and methodologies, they cannot be used to compare the considered standards between each other.

However, these numbers show us how far practical systems are from the theoretical limit derived using the Shannon formula for infinite bandwidth and path loss of 83 dB, i.e., 0.55 $pJ/b$.

Similarly to energy consumption, the latency introduced by wireless links depends on multiple factors. The time of flight between the signal source and its destination is proportional to the distance and inversely proportional to the speed of light. It is negligible (below a few μs) for a typical wireless link up to a few km. Transmission time is more important. It is proportional to the payload (number of bits) and inversely proportional to the link throughput. However, there are other factors increasing latency, like the time needed by the Automatic Repeat reQuest (ARQ) procedure that is dependent on an internal characteristic of the utilized wireless standard. Next, the transmitted packet can be subject to a random delay caused by the utilized Medium Access Control (MAC) scheme, its configuration, and the number of users competing for a wireless medium at the same time. Finally, some random phenomena in the wireless channel, e.g., fast fading, can cause an outage of the link, increasing the transmission latency. While all these factors are difficult to be accurately described by a single model, measurement-based models are of high potential. Measurements of a Round Trip Time between LTE/5G UE and the Base Station from [4] are presented in Table I (mean values). This is a value for a short packet dominated by the MAC and ARQ procedures. Most importantly, the value is significantly below the limit of 4 ms specified for eMBB in 5G and above the limit of 0.5 ms for URLLC [4]. For WiFi networks, the dominating factor will be the MAC procedure requiring all transmitting devices to compete for spectral resources. As shown in [6] the induced delay is quasi-exponentially distributed for a single packet with a mean delay ranging from a few to a few hundred ms.

### B. Wired part of the network

Wired connections have always played a key role in the development of the Internet. They are essential in the modern Internet from its access (e.g., Passive Optical Networks (PONs)) to its core (e.g., Elastic Optical Networks (EONs)). Taking the perspective of 2C networks (Fig. 1), wired connections of end devices to the edge/fog nodes (e.g., a laptop connected to a router using an Ethernet cable) are relatively rarely used nowadays due to their limited flexibility, even though their energy efficiency is higher than the efficiency of wireless links. Wired links are mainly used for interconnecting edge/fog nodes, as well as for connecting the edge/fog tier with the cloud. Gigabit Ethernet realized on copper cables is usually sufficient for interconnecting edge/fog nodes. However, Wavelength Division Multiplexing (WDM) links realized on optical cables are more suitable for these interconnections due to their higher bandwidth.

Performing computations in the cloud is more effective than performing them in the edge/fog due to the parameters of computing devices (high computational power, effective cooling, etc.). However, transporting computation tasks to the cloud as well as the computation outcomes back to the end-user can be time-consuming due to the physical distance between the fog and the cloud [6]. The few IP routers that the computation task needs to travel through may also influence experienced jitter. On the other hand, little additional energy is needed for the transportation of the tasks to the cloud due to the low dependence of power consumption of core IP routers and optical devices on load [7]. This is indicated in Table I, where energy efficiency is based on the extra power needed for sending packets with respect to idle power. Induced latency is mainly determined by the propagation time of the optical signal reaching the highest values for submarine cables. Dense WDM and EONs are used in the core of the Internet to realize the connection between the cloud and the edge/fog nodes.

*C. Computing*

The purpose of green computing is to increase energy efficiency over the course of a computing device's lifetime. Standard methods for achieving this energy efficiency are data-center design, software and deployment optimization, power management, optimized cloud computing, and edge/fog computing. Algorithmic efficiency, optimized computing resource allocation, machine virtualization, Dynamic Voltage and Frequency Scaling (DVFS), sleeping modes, and the use of terminal servers are in place to optimize the software and deployment of the computing machines. These measures are taken to reduce the energy consumption resulting from the computers themselves and their air-conditioning and ventilation systems.

The performance-per-watt efficiency of the top 500 most energy-efficient supercomputers (Green500) [8] is continuously increasing with the top 2 supercomputers recently reaching values over 60 GFLOPS/W. However, as shown in [9], PCs are a plausible, energy-efficient option for the execution of non-complex tasks. Measurements of performance rates of tasks' execution and power efficiency (in GFLOPs/Watt) of five PCs are presented in [9]. In Table I, key performance metrics of selected supercomputers and a PC are compared. Energy efficiency is provided in pJ/s for consistency reasons, assuming 71-220 GFlop/B as a range of aggregate arithmetic intensities [10]. The *Henri* supercomputer (no. 1 on the Green500 list) has the best energy efficiency in Joules per bit, while the *Frontier* supercomputer (ranked no. 1 on the list of best performance supercomputers Top500) has worse energy efficiency but also lower power consumption and more than 500 times higher processing speed in Flop/s. Interestingly, an exemplary PC (Asus Expertbook, Core i7-1165G7 2.8 GHz [9]) has better energy efficiency than supercomputer *Cumulus* which ranked 106 on Green500 (see Table I). Naturally, the processing speed is not as high for this PC as for supercomputers, but this example shows that when less computationally demanding tasks are to be executed, some less powerful but more energy-efficient machines are a viable option. Additionally, they are supposed to be localized closer to the end devices at the edge of a network.

Table I. Energy and latency consumption.

| Wireless Link | Bandwidth [MHz] | TX eff. [pJ/b] | RX eff. [pJ/b] | Latency [ms] |
|---|---|---|---|---|
| Shannon limit | ∞ | 0.55 | 0 | - |
| Wi-Fi link [3] | 20 | 4.5e4 | 3.9e4 | $1 - 1000$ |
| LTE UE DL [4] | 20 | - | ~1.7e6 | RTT: 2.6 |
| 5G UE DL [4] | 100 | - | ~4e5 | RTT: 2.2 |
| LTE BS DL [5] | 10 | 4.5e7 | - | - |

| Wired Link | Capacity [Gb/s] | Active power [W] | Eff. [pJ/b] | Latency [ms] |
|---|---|---|---|---|
| 1G EPON gateway [11] | 1 | 3.3 | 300 | $0,5e-5 - 0,5$ |
| 10/10G GPON gateway [11] | 10 | 5.5 | 200 | as above |
| Juniper T1600 core router [7] | 640 | 6572 | 1030 | $0,01 - 27$ |

| (Super) computer | Perf. [TFlop/s] | Cores | Power [kW] | Eff. [pJ/b] |
|---|---|---|---|---|
| Henri (#1 Green500) [8] | 2038 | 5920 | 31 | $136 - 422$ |
| Frontier (#1 Top500)[8] | 1102e3 | 8730112 | 21100 | $170 - 527$ |
| ASUS laptop Expertbook B9400CEA [9] | 0.148 | 4 | 0.03347 | $2000 - 6199$ |
| Cumulus (#106 Green500) [8] | 2271.38 | 50176 | 530 | $2069 - 6410$ |

## III. Optimization of Energy Consumption in 2C Fog Networks

Both communication and computing introduce latency and energy costs in the network, as discussed in Sec. II. On the one hand, an individual device from the things tier (e.g., a smartphone) is usually battery-powered and has limited resources compared to fog nodes and cloud nodes. Therefore, it can aim at the optimization of its energy consumption/utility, disregarding the costs in the higher tiers of the network treating it as a service provider. On the other hand, from the point of view of a network operator, optimizing the total energy costs in the network (while maintaining the required QoS, e.g., latency) could be the goal. In the first case, the decision boils down to whether the costs related to transmitting the task outweigh those caused by processing it locally by the device. In the second case, the fog network can distribute resources (networking and computing) – choose which nodes should process the tasks offloaded by the users, and how it should be done, e.g., using what CPU frequency. This optimization can be done after the tasks are sent by the users to their access point. This scenario is examined in Sec. III-A. The optimization could also be carried out considering Radio Access Network (RAN). Then, the transmission from end devices to the fog is optimized jointly with the processing of tasks within fog and cloud nodes. This scenario is shown in Sec. III-B.

Finally, the optimization can take into account that many fog applications are for periodic requests. In this case, it is both the timeliness and accuracy that specify if the QoS required by a given application is met. For this purpose, the AoI metric is currently used to optimize communications [12], though it can be easily extended to take 2C into account. AoI is defined as the time elapsed since the latest request (whose computation result reached the destination) has been generated. The request generation rate in the source influences the rate required in links, the number of computations to be carried in the fog, and the possible queuing of requests as such influencing the AoI. At the same time, energy utilization is impacted. Variable packet generation rate at source in order to minimize energy consumption while maintaining required AoI is considered in Sec. III-C. Observe that while Sec. III-A proposes to optimize communications in the wired part of the network and Sec. III-B extends it with RAN optimization, Sec. III-C proposes another degree of freedom by considering the request source utilizing its periodic behavior. All considered optimization scenarios are compared in Table II. All of these can be applied to various time-critical applications thanks to tight latency or AoI constraints utilized.

These optimization problems can be sophisticated. Apart from the main decision variables which are binary (whether or not to process/offload a task) or integer (where to send/process it), there are other parameters such as clock frequency and transmit power (i.e., continuous variables) which make it nontrivial to find the optimum. In the following sections, illustrative results of simulations are shown and discussed.

Table II. Summary of optimization scenarios.

| Scenario | Energy costs | Constraints | Variables |
|---|---|---|---|
| Sec. III-A | Spent by fog and cloud nodes | Latency | Main: offloading decision (which node computes) Aux: CPU frequency, transmission rate |
| Sec. III-B | Spent by end devices and fog and cloud nodes | Latency | Main: offloading decisions (which nodes to transmit, which node computes) Aux: CPU frequency, transmission rate |
| Sec. III-C | Spent by end devices and fog node | AoI | Main: operating frequency, transmission rate |

### A. Optimization within the fog and cloud tiers

Let us assume the following scenario: there are 10 interconnected fog nodes with a connection to the cloud through the Internet. End devices wirelessly send computational requests to these nodes. The requests are characterized by size, arithmetic intensity (required number of operations relative to the size), and maximum tolerated delay. For such a network, the optimization problem can be defined as the minimization of energy consumption spent on computing and transmitting these requests while satisfying their delay constraints as in [6]. It is achieved by distributing requests to nodes for computations and adjusting the CPU frequency of nodes through DVFS.

Fig. 2 plots the cumulative distribution function of energy costs related to offloading requests considering 3 task allocation strategies. The results are achieved through computer simulations according to the model and optimization shown in [6]. The blue and magenta lines represent an approach in which arriving requests are computed in the same node to which they were transmitted by the end device – there is no inter-fog or fog-to-cloud transmission of requests. There is inter-fog and fog-to-cloud transmission for the red line (full optimization). Blue and red lines show results in which fog nodes are computing at optimal frequencies, while in the magenta solution computations are performed at the maximal available CPU frequency.

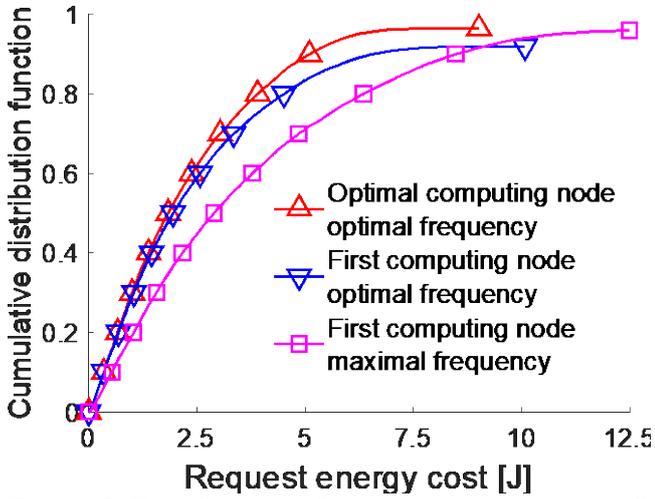

Figure 2. Distribution of energy costs spent on offloaded requests

One can see that the red line (full optimization) is further left than the blue one (it achieves lower energy costs) and also further up (it is able to successfully process more requests). Computing at maximal frequencies (magenta line) induces significantly higher energy costs, while successfully processing a similar number of requests as the fully optimized solution (red). By comparing the median request energy, it is visible that around 30% of energy can be saved if instead of computing requests in the closest fog node with the highest CPU frequency (the magenta line), optimal allocation to computing nodes along with CPU frequency adjustment is carried out (the red line).

*B. Optimization between wireless network, fog, and cloud tiers*

In the previous section, the optimization began upon the appearance of requests in the fog nodes. Let us consider the same network and requests as in Sec. III-A, but add a decision point "to which fog node should this end device wirelessly send this request". It also adds a new level of complexity to the existing optimization problem of choosing the optimal wireless transmission rate.

Fig. 3 shows the results of optimization for this scenario. Here, the results are generated through simulation according to the model and optimization shown in [13]. Medians of energy costs spent on offloaded requests are plotted after a size parameter sweep. The red line shows the best results achieved by choosing the optimal computing node, optimal transmission path, and rate, as well as the optimal CPU frequency. The blue line shows a scenario where all requests are processed in the fog nodes closest to the corresponding end devices. The magenta line shows the same scenario as the blue one, except fog nodes work at the maximal available CPU frequency. The black line corresponds to an optimization where each request is computed in the fog node collocated with the wireless access point to which the request has been originally transmitted.

Fig. 3 shows that the optimal distribution of requests to nodes achieves the lowest energy consumption. The differences between plotted values increase as the sizes of requests increase. Baseline solutions shown in blue, black, and magenta "abruptly end" before reaching the right end of the plot. It corresponds to the fact that less than 50% of all requests were successfully processed within the tolerated delay. The difference between results obtained by different solutions varies with the size of the request. At 2 MB the optimal (red) solution saves 49.5% energy when compared with the solution shown in magenta, 23.5% when compared with blue, and 4.6% when compared with black. At 6 MB these savings change to 44.6%, 31.8%, and 12.2% respectively.

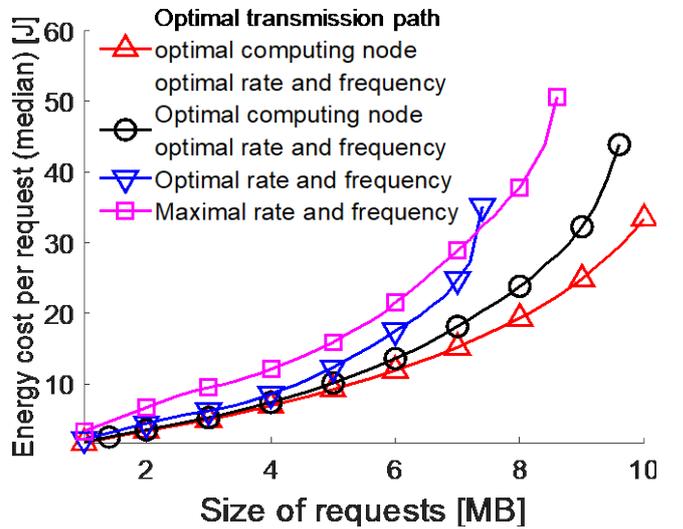

Figure 3. Median of energy costs spent on offloaded requests as a function of request size

*C. Optimization of requests generation rate considering the AoI for 2C*

Let us consider an end device that generates a certain number of requests per unit of time. Each request must be transmitted to a proper fog node and processed therein to obtain useful information. If the resources at any stage (wireless network, wired network, computing nodes) are not available, the request is queued in a First-In-First-Out (FIFO) buffer. The request is sent to the base station over a single time slot with the transmission power minimizing energy consumption. Moreover, the CPU frequency in the fog node is optimized in order to minimize energy consumption. The mean consumed power, as well as the mean AoI versus request generation rate are plotted in Fig. 4. It can be observed that initially (up to 0.9 packets/ms) an increase in the requests generation rate results in a decrease in AoI. However, AoI starts to increase next as a result of computational or communications resources being exhausted, requiring requests to be queued. On the other hand, it is visible that an increase in the request generation rate results in a rise in power consumption. However, when the 2C resources become fully utilized, the mean required power reaches a defined plateau. It is reasonable as real world devices reach the maximum power consumption when fully utilized. Depending on the required AoI for a given application, an optimal request generation rate can be configured, such that the mean power consumption is minimized. This shows that source optimization should be considered together with wireless transmission, wired transmission, and fog nodes in order to maximize 2C efficiency while obeying the latency constraints.

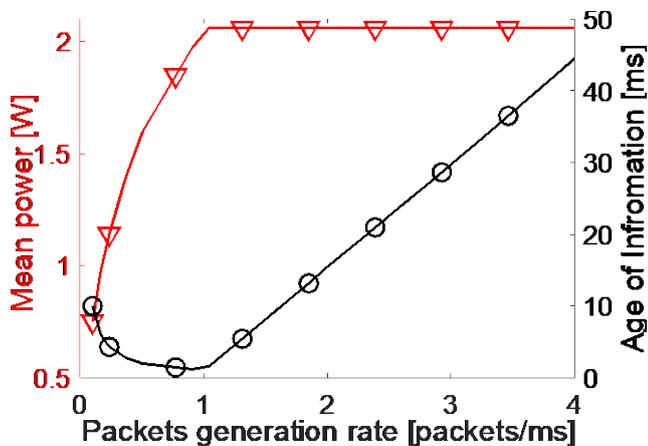

Figure 4. Mean power consumption and AoI versus the request generation rate.

**IV. AI for 2C Energy Management**

The previous section has shown that increasing the number of degrees of freedom in optimization enables energy efficiency to increase without deterioration of the Quality of Service. However, global optimization considering numerous factors (e.g., fog nodes CPUs' frequency or the wireless access point to be used) can be problematic. First, it requires a global view of the considered system including all power consumption models and potential delays introduced by the considered allocation. Second, the optimization should be done without delays influencing internal networks of various service providers, e.g., wireless radio access networks or cloud computing centers.

Artificial Intelligence (AI) and Machine Learning (ML) techniques can be employed. These can be used to learn the power consumption models or latency models while observing real-time network parameters and the induced energy consumption using, e.g., some reinforced learning approach. While the latency can vary randomly as a result of, e.g., random fading in the wireless channel, its distribution can be learned based on the available network parameters. Another perspective is to directly employ AI to select task allocation. This can use, e.g., Deep Reinforcement Learning, as proposed in [14]. Previously tested allocation strategy will be assessed and after some operation time, it should converge to an optimal or close to the optimal solution. Additionally, utilization of such a scheme allows the allocation to adapt, e.g., to changing conditions, traffic on each link, or changes of fog nodes.

Finally, ML can be used to design tasks allocation policies working independently, e.g., in each fog node. In this case, it can be beneficial to utilize some clustering schemes. The incoming computation requests can be clustered according to their properties, e.g., number of calculations required, number of bits to be transmitted, or latency constraint, using the k-means algorithm [15]. Each cluster can be assigned a different strategy, e.g., local computation or offloading to the cloud. The strategies for each node and cluster can be obtained using, e.g., reinforced learning. It is also worth noting that the costs of training networks can be non-negligible. Ideally, costs spent on training and optimization should be included when examining the efficiency of ML-based and non-ML-based solutions. It is an interesting topic for future work.

**V. Conclusions**

Energy efficiency of fog computing becomes a major problem, especially for many time-critical applications. We have shown in Sec. III that it is possible to reduce energy consumption through proper coordination of communicating and computing resource allocation. The energy consumption can be further improved by proper source management depending on the current 2C network status. While centralized optimization is difficult to implement, we believe that distributed, AI-driven 2C management algorithms can achieve energy efficiency close to the global maximum.

**Acknowledgment**

The work has been funded by the National Science Centre in Poland within the FitNets project no. 2021/41/N/ST7/03941 on "Fresh and Green Cellular IoT Edge Computing Networks - FitNets."

**Literature**

[1] Global e-Sustainability Initiative and Deloitte, "Digital with Purpose: Delivering a

[1] SMARTer2030", 2019, https://gesi.org/research/gesi-digital-with-purpose-full-report, accessed: 2022-12-13.
[2] M. Chiang, B.h Balasubramanian, and F. Bonomi (Eds.), *Fog for 5G and IoT*, Information and Communication Technology Series, Wiley, 1 edition, 2017.
[3] P. Kryszkiewicz, A. Kliks, Ł. Kułacz, and B. Bossy, "Stochastic Power Consumption Model of Wireless Transceivers," *Sensors* 20, no. 17: 4704, 2020.
[4] D. Xu et al. "Understanding operational 5G: A first measurement study on its coverage, performance and energy consumption," *Proc. SIGCOMM,* 2020.
[5] G. Auer et al., "How much energy is needed to run a wireless network?," in IEEE Wireless Communications, vol. 18, no. 5, pp. 40-49, 2011.
[6] B. Kopras, B. Bossy, F. Idzikowski, P. Kryszkiewicz, and H. Bogucka, "Task Allocation for Energy Optimization in Fog Computing Networks with Latency Constraints," in IEEE Transactions on Communications, vol. 70, no. 12, pp. 8229–8243, 2022.
[7] W. Van Heddeghem, F. Idzikowski, W. Vereecken, D. Colle, M. Pickavet, and P. Demeester, "Power consumption modeling in optical multilayer networks," Photonic Network Communications, vol. 24, no. 2, pp. 86–102, 2012.
[8] Green500 Release November 2022, https://www.top500.org/lists/green500/2022/11/, last accessed on 2022-12-13.
[9] B. Prieto, J. J. Escobar, J. C. Gómez-López, A. F. Díaz, and T. Lampert, "Energy Efficiency of Personal Computers: A Comparative Analysis," Sustainability, vol. 14, no. 19, p. 12829, 2022.
[10] J. Kosaian and K. V. Rashmi, "Arithmetic-Intensity-Guided Fault Tolerance for Neural Network Inference on GPUs," SC21: International Conference for High Performance Computing, Networking, Storage and Analysis, 2021.
[11] P. Bertoldi and A. Lejeune, "Code of conduct on energy consumption of broadband equipment: Version 8.0," 2021.
[12] L. Zhang, L. Yan, Y. Pang and Y. Fang, "FRESH: FReshness-Aware Energy-Efficient ScHeduler for Cellular IoT Systems," ICC 2019 - 2019 IEEE International Conference on Communications (ICC), 2019.
[13] B. Kopras, B. Bossy, F. Idzikowski, P. Kryszkiewicz and H. Bogucka "Communication and Computing Task Allocation for Energy-Efficient Fog Networks," in Sensors. 2023; 23(2):997.
[14] H. Zhou, K. Jiang, X. Liu, X. Li, and V. C. M. Leung, "Deep Reinforcement Learning for Energy-Efficient Computation Offloading in Mobile-Edge Computing," in *IEEE Internet of Things Journal*, vol. 9, no. 2, pp. 1517-1530, 2022.
[15] A. Alnoman, "Machine Learning-Based Task Clustering for Enhanced Virtual Machine Utilization in Edge Computing", 2020 IEEE Canadian Conference on Electrical and Computer Engineering, 2020, London, Canada.

**Hanna Bogucka** is a professor at the Institute of Radiocommunications (IR) at Poznan University of Technology (PUT). She is involved in research in wireless, cognitive, and green communications. She is a member of the Polish Academy of Sciences. She also serves as the Member-at-Large of IEEE Communications Society Board of Governors and Europe Regional Chair in IEEE ComSoc Fog/Edge Industry Community.

**Bartosz Kopras** received the B.Sc. degree and the M.SC. degree in telecommunications from PUT in 2019 and 2020, respectively. He is a Ph.D. student at the IR at PUT. His main field of interest is fog networks.

**Filip Idzikowski** received the M.S. degree in telecommunication engineering from the PUT, Poland, and Dublin City University, Ireland, and the Ph.D. degree from the Technical University of Berlin, Germany. He is currently an Assistant Professor with PUT. He regularly serves as a member of various committees on several conferences, including IEEE ICC and IEEE Globecom.

**Bartosz Bossy** received the M.Sc. degree and the Ph.D. degree in telecommunications from PUT in 2015 and 2022, respectively. He is a Research Assistant with the IR. He has been involved in a number of national and international projects. His research interests include green communications, energy-efficient resource allocation, fog networks and optimization.

**Paweł Kryszkiewicz** is currently an Assistant Professor with the IR, PUT. He has been involved in a number of national and international research projects. His main fields of interest are multicarrier system design, green communications, Dynamic Spectrum Access, and Massive MIMO systems.